\documentclass[aps,prl,reprint,showpacs,floatfix,superscriptaddress,nofootinbib]{revtex4-2}

\usepackage[T1]{fontenc}
\usepackage[utf8]{inputenc}
\usepackage[english]{babel}
\usepackage[normalem]{ulem}

\usepackage{amsmath}
\usepackage{mathrsfs} 
\usepackage{lipsum}
\usepackage{amsfonts}
\usepackage{amssymb}
\usepackage{amsthm}
\usepackage{mathtools}
\usepackage{latexsym}
\usepackage{bm}
\usepackage{relsize}
\usepackage{braket}
\usepackage{slashed}
\usepackage{empheq}
\usepackage[makeroom]{cancel}
\usepackage{xfrac} 
\usepackage{upgreek}

\usepackage{graphicx}
\usepackage{tabularx}
\usepackage{multirow}
\usepackage{floatrow}
\usepackage{booktabs}

\usepackage{pgf}

\usepackage{xspace}
\usepackage{hyperref} 
\usepackage[most]{tcolorbox}
\usepackage{comment}

\usepackage{tikz-feynman}
\usepackage{tikz}	
\tikzfeynmanset{compat=1.0.0}

\allowdisplaybreaks

\usepackage{colortbl}
\definecolor{summersky}{cmyk}{0.71,0.33,0,0.5}
\definecolor{flamingo}{cmyk}{0,0.51,0.71,0.5}
\definecolor{rp}{cmyk}{0.2, 1, 0.6, 0}
\definecolor{pacificblue}{cmyk}{0.95,0.3,0, 0.5}
\definecolor{gray60}{cmyk}{0.4,0.4,0,0.8}

\newcommand{\ie}{\textsl{i.e.}~}

\newcommand{\dd}{\mathrm{d}}
\newcommand{\ee}{e}

\newcommand{\sss}[1]{{\scriptscriptstyle{#1}}}
\newcommand{\boldmathsymbol}[1]{{\ensuremath{\boldsymbol{#1}}}}

\newcommand{\uS}{\mathrm{S}}

\newcommand{\usssS}{\sss{\uS}}

\newcommand{\nS}{n_\usssS}
\newcommand{\AS}{A_\usssS}

\newcommand{\bmk}{\boldmathsymbol{k}}

\newcommand{\bmx}{\boldmathsymbol{x}}

\newcommand{\cs}{c_\text{\tiny{S}}}

\newcommand{\beq}{\begin{equation}}
\newcommand{\eeq}{\end{equation}}
\newcommand{\bea}{\begin{equation}\begin{aligned}}
\newcommand{\eea}{\end{aligned}\end{equation}}

\newlength{\wsingfig}
\newlength{\wdblefig}
\newlength{\wquadfig}
\newlength{\wtriplefig}
\newcolumntype{P}[1]{>{\centering\arraybackslash}p{#1}}

\begin{document}

\title{
Universal CMB Phase Coherence from Single-Field Inflation}

\author{Siméon Vareilles}
\affiliation{Aix Marseille Univ, Universit\'e de Toulon, CNRS, CPT, Marseille, France}

\author{Thomas Colas}
\affiliation{Department of Applied Mathematics and Theoretical Physics, University of Cambridge, Wilberforce Road, Cambridge, CB3 0WA, UK}

\author{Julien Grain}
\affiliation{Universit\'e Paris-Saclay, CNRS, Institut d'Astrophysique Spatiale, 91405, Orsay, France}

\author{Federico Piazza}
\affiliation{Aix Marseille Univ, Universit\'e de Toulon, CNRS, CPT, Marseille, France}

\author{Vincent Vennin}
\affiliation{Laboratoire de Physique de l'\'Ecole Normale Sup\'erieure, CNRS, ENS, Universit\'e PSL, Sorbonne Universit\'e, Universit\'e Paris Cit\'e, F-75005 Paris, France}

\begin{abstract}
The phase coherence of primordial perturbations shapes the acoustic peaks of the CMB and is regarded as a key signature of inflation. Inflationary scenarios at different energy scales, thus with different durations, are commonly expected to leave different levels of coherence. We show this expectation fails: because the curvature perturbation is conserved on super-Hubble scales, the coherence at recombination inherits a value fixed by the observed primordial spectrum alone. This holds irrespective of the inflationary energy scale and the subsequent expansion history.
\end{abstract}

\maketitle

\paragraph*{Introduction.---}
%

Among the many successes of inflation \cite{Starobinsky:1979ty,Guth:1980zm, Starobinsky:1980te,Sato:1980yn, Linde:1981mu, Mukhanov:1981xt, Guth:1982ec, Albrecht:1982wi, Starobinsky:1982ee, Hawking:1982cz, Linde:1983gd, Bardeen:1983qw, Mukhanov:1988jd}, one prediction stands out over competing scenarios for the origin of the primordial cosmological perturbations: phase coherence \cite{Hu:1995kot, Hu:1996yt, Dodelson:2003ip}. 
This feature underlies the acoustic peaks observed in the temperature and polarization anisotropies of the Cosmic Microwave Background (CMB), and the Baryon Acoustic Oscillation (BAO) feature in the large-scale structure of the universe. In particular, the alternating sign pattern of the temperature--polarization (TE) cross-correlation, including its anticorrelation on one-degree scales, is difficult to reconcile with incoherent causal sources such as topological defects and provides strong evidence for a coherent, super-horizon primordial phase distribution~\cite{Dodelson:2003ip, WMAP:2003ggs,Planck:2018vyg}.

Phase coherence is most simply characterised at the level of the Mukhanov-Sasaki variable in Fourier space $v_{\bmk}$, which throughout the cosmological history obeys the equation of a harmonic oscillator with time-dependent frequency $\omega_k$ (see eqs.~\eqref{eq:action_fourier}--\eqref{eq:fourier_modes} below). Its solutions admit the WKB form
\begin{equation}\label{eq:wkb}
  v_\bmk = \frac{A_\bmk}{\sqrt{2W_k(\eta)}}\,
  \cos\!\left[\int_0^{\eta}\dd\tilde\eta\,W_k(\tilde\eta)
  + \theta_\bmk\right],
\end{equation}
where $W_k=\omega_k+\ldots=\cs k+\ldots$ (dots denoting higher-order WKB corrections) is fixed by the background dynamics, and the initial conditions determine $A_\bmk$ and $\theta_\bmk$. It is then natural to define the phase at any time as the angle~\cite{Susskind:1964zz}
\begin{equation}\label{eq:phase_definition}
\theta_\bmk\equiv\arctan\!\left(\frac{v_\bmk^{\prime}}{\cs k\,v_\bmk}\right)\!\!\pmod\pi\,,
\end{equation}
which on sub-horizon scales  (\ie at leading order in WKB) differs from the phase involved in eq.~\eqref{eq:wkb} by the deterministic quantity $\int_0^\eta W_k$, and
hence shares its statistical dispersion when the direction of the wavenumber $\bmk/k$ is varied. Of course one could work with more directly observable variables, such as $\delta T/T$. Adiabatic initial conditions, however, leave a single degree of freedom at each $\bmk$, and the variables describing it differ only by background-dependent rescalings: for an arbitrary $\zeta = f(\eta)\,v$ with $f$ a background quantity,
$\zeta_\bmk'/\zeta_\bmk = v_\bmk'/v_\bmk + f'/f$, and the $\bmk$-independent
shift $f'/f$ moves the mean phase while leaving its dispersion unchanged.   

Observations are compatible with $A_\bmk$ being stochastic variables,
with the corresponding curvature power spectrum famously of order $\mathcal{P}_{\zeta}\sim 10^{-9}$~\cite{COBE:1992syq}, while the fluctuations in the phases $\theta_\bmk$ have an extraordinarily small (and, to this day, unresolved) dispersion~\cite{Albrecht:1995bg,Magueijo:1995xj,
Dodelson:2003ip,Planck:2018vyg}. This, in essence, is phase coherence, and is responsible for the fixed positions of the acoustic peaks in the CMB.
Phase coherence is equivalently attributed to the conservation of the curvature perturbation $\zeta = v/z$ on super-Hubble scales \cite{Maldacena:2002vr, Weinberg:2003sw, Pajer:2017hmb}, the suppression of the decaying mode \cite{Polarski:1995jg, Kiefer:2008ku, Sudarsky:2009za}, or the large squeezing of the quantum state generated during inflation \cite{Grishchuk:1990bj, Albrecht:1992kf, Grishchuk:1992tw,Albrecht:1996dq, Grain:2019vnq, Colas:2021llj, Piotrak:2025zhy}. These all imply ``negligible initial velocity'' of the adiabatic curvature perturbation when it crosses back inside the Hubble radius, thus $\Delta\theta_\bmk\simeq  0$.

Coherence is widely expected to be determined primarily by inflationary squeezing, which grows with the time a mode spends outside the Hubble radius, and thus retains a memory of the inflationary energy scale and of the details of reheating. In this Letter, we show that this intuition is incomplete. Keeping track of the quantum state of the perturbations through out inflation and the subsequent eras, we find that post-inflationary evolution contributes as much to the final coherence as inflation itself. In particular, the squeezing generated at the transition to radiation domination compensates the energy-scale dependence accumulated during inflation, regardless of the reheating details, leading to a universal phase coherence. The final prediction is fixed entirely by the observed amplitude of temperature fluctuations and is extraordinarily small: for slow-roll inflation the dispersion is of order $\Delta\theta_{k_\star}\simeq e^{-117}$ on CMB scales. Since this conclusion only requires the curvature perturbation to be conserved on super-Hubble scales, any departure from that prediction
would point to a breakdown of this condition.\\

\begin{figure*}[t!]
  \centering
  \includegraphics[width=0.368\textwidth]{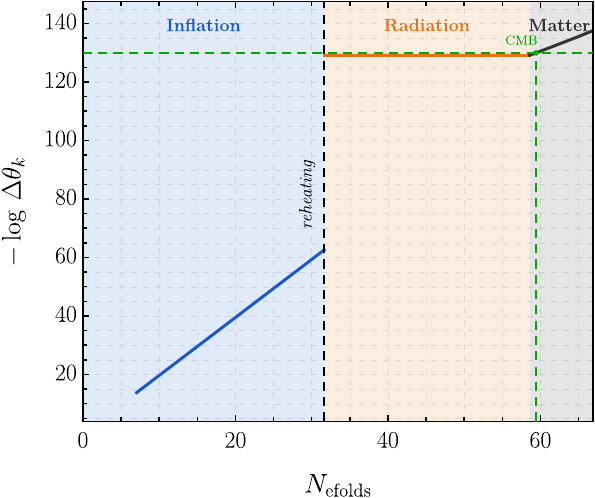}
  \hspace{-0.6em} 
  \includegraphics[width=0.618\textwidth]{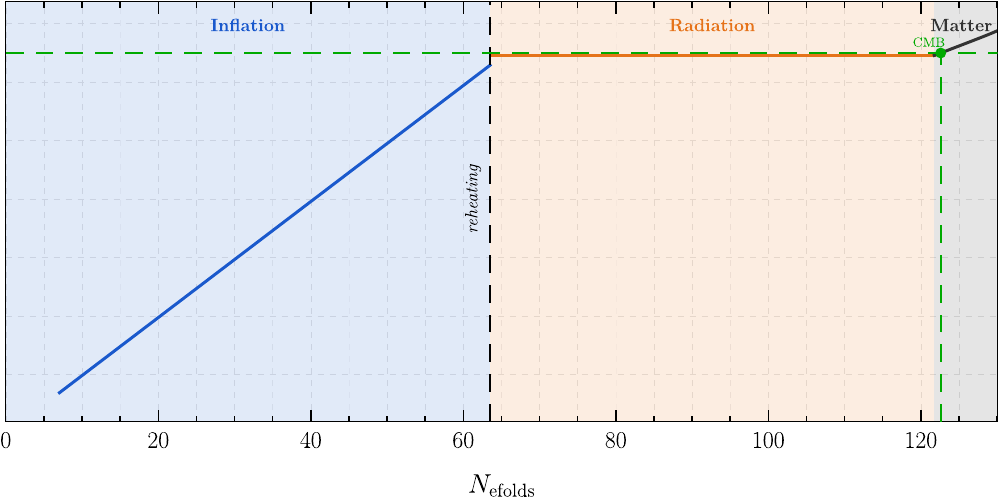}
  \caption{Evolution of the logarithmic phase variance $-\log\Delta\theta_k$ for a fixed super-horizon mode, $k=10^{-4}\,\mathrm{Mpc}^{-1}$, chosen small enough that it remains super-Hubble until well after recombination, as a function of the number of $e$-folds for EW (\textit{left}) and GUT (\textit{right}) inflationary scenarios. The three linear regimes correspond, from left to right, to inflation, radiation domination, and matter domination, while the jumps encode the matching conditions. Although the two scenarios exhibit vastly different coherence levels at the end of inflation, they converge to the same phase coherence on CMB scales: the jump at reheating compensates the coherence accumulated during inflation, cancelling the dependence on the inflationary energy scale.}\label{fig:mainfig}
\end{figure*}

\paragraph*{Phase coherence from squeezing.---}

For a flat-FLRW universe $\dd s^2 = a^2(\eta) \left(- \dd \eta^2 + \dd\bmx^2\right)$, filled either by a perfect barotropic fluid or by a scalar field, with equation of state parameter $w$ and speed of sound $\cs $, the quadratic action for the {Mukhanov-Sasaki variable} $v(\eta,\mathbf{x})=\int \dd^3\bmk\, v_\bmk(\eta)\,e^{i\bmk\cdot\mathbf{x}}$ reads in Fourier space
\begin{equation}
    S_{v}^{(2)}=\frac12\int \dd\eta \int \dd^3\bmk\,
    \left(v_{\bmk}'v_{-\bmk}'-\omega_k^2\,v_{\bmk}v_{-\bmk}\right)\,,
    \label{eq:action_fourier}
\end{equation}
where $z^2 \equiv  3M_{\text{Pl}}^2(1+w)\, a^2/\cs ^2$. Primes denote derivatives with respect to conformal time, and the time-dependent frequency is $\omega_k^2(\eta) \equiv \cs ^2 k^2 - z''/z$. 
The associated Fourier mode functions $v_k$ satisfy
\begin{equation}
    v_k''+\omega_k^2\,v_k=0\,.
    \label{eq:fourier_modes}
\end{equation}
Since $\omega_k$ depends only on the modulus $k\equiv|\bmk|$ of the wavevector, so do the mode functions $v_k$ and all spectra below, provided initial conditions are set in the isotropic Bunch-Davies vacuum. The field $v(\eta,\mathbf{x})$ being real, this ties $\bmk$ to $-\bmk$ ($\hat{v}_\bmk^{\dagger}=\hat{v}_{-\bmk}$), so the Hamiltonian decouples into independent two-mode subsystems. Upon quantization, each is a parametric oscillator with canonical pair $\{\hat{v}_\bmk, \hat{\pi}_\bmk=\hat{v}^{\prime}_\bmk\}$ satisfying $[\hat{v}_\bmk,\hat{\pi}_{\bmk'}]=i\,\delta^3(\bmk+\bmk')$. The Hamiltonian being quadratic, the Schr\"odinger equation is solved by a Gaussian wavefunctional $\Psi_\bmk[v]\propto\exp\!\left[-\tfrac12\,\Omega_k(\eta)\,v_\bmk v_{-\bmk}\right]$, fully specified by the power spectra,
\begin{equation}
\begin{aligned}
    \langle \hat v_{\bmk}\hat v_{\bmk'}\rangle
        &= \delta^3(\bmk+\bmk')\,P_{vv}\,, & P_{vv}&=|v_k(\eta)|^2\,,\\
    \tfrac12\langle\{\hat v_{\bmk},\hat\pi_{\bmk'}\}\rangle
        &= \delta^3(\bmk+\bmk')\,P_{vp}\,, & P_{vp}&=\mathrm{Re}\!\left[v_k(\eta)\,v_k^{\prime *}(\eta)\right]\,,\\
    \langle \hat\pi_{\bmk}\hat\pi_{\bmk'}\rangle
        &= \delta^3(\bmk+\bmk')\,P_{pp}\,, & P_{pp}&=|v_k'(\eta)|^2\,,
\end{aligned}
\label{eq:powers_pectra}
\end{equation}
written in terms of the solutions of Eq.~\eqref{eq:fourier_modes}. Unitary evolution keeps the state pure, saturating the uncertainty bound, $P_{vv}P_{pp}-P_{vp}^2=1/4$, such that $\Omega_k=1/(2P_{vv}) - i\,P_{vp}/P_{vv}$.  This unitary evolution is a Bogoliubov transformation mixing the creation and annihilation operators, squeezing the initial adiabatic vacuum into a two-mode squeezed state \cite{Grishchuk:1990bj,Albrecht:1992kf,Polarski:1995jg,Grain:2019vnq}. The Gaussian state can therefore equivalently be entirely characterized by a squeezing amplitude $r_k$ and a squeezing angle $\phi_k$. In terms of the power spectra, these parameters read~\cite{Grain:2019vnq}
\begin{equation}
\begin{aligned}
  \cosh(2r_k)&= \cs  k\,P_{vv}(k)+ (\cs  k)^{-1}\,P_{pp}(k)\,,\\
  \tan(2\phi_k)&= \frac{2\,P_{vp}(k)}{\cs  k\,P_{vv}(k)- (\cs  k)^{-1}\,P_{pp}(k)}\,.
\end{aligned}
\label{eq:squeezing}
\end{equation}

The squeezing parameters, however, are not symplectic invariants: they depend on the choice of canonical pair. 
What we ultimately seek is the \emph{phase} of the perturbations, which provides a natural measure of their coherence at last scattering and hence of the CMB acoustic peaks. The quantity defined in Eq.~\eqref{eq:phase_definition} captures this phase, reducing to the leading WKB phase of the Mukhanov--Sasaki variable on sub-sonic scales (see Supplemental Material).\footnote{In practice, the squeezing angle $\phi_k$ of the state is subtracted
from eq.~\eqref{eq:phase_definition}, so that the resulting distribution
is centred on zero and $\langle\theta_\bmk\rangle=0$ by construction.\label{footnote:remove:phik}} For each Fourier mode, $\theta_{\bmk}\in\left]-\tfrac{\pi}{2}\,,\tfrac{\pi}{2}\right[$ is a stochastic variable whose spread across realizations characterizes the loss of phase coherence. Its dispersion, $(\Delta\theta_k)^2=\langle\,\theta_{\bmk}^2\,\rangle$, averaged over $\bmk$ at fixed amplitude $k$, therefore quantifies phase coherence.

To investigate the statistical properties of the phase~\eqref{eq:phase_definition}, we use the fact that the pair $(v_\bmk, v^{\prime}_\bmk/(\cs k))$ is a zero-mean bivariate Gaussian entirely specified by its covariance matrix
\begin{equation}
    \begin{aligned}
    \Sigma&=\begin{pmatrix}P_{vv} & P_{vp}/(\cs k) \\[2pt]
P_{vp}/(\cs k) & P_{pp}/(\cs ^2k^2)\end{pmatrix},
\end{aligned}
\end{equation}
with $\det\Sigma=1/(4\cs ^2k^2)>0$ and the joint Gaussian bivariate probability density is given by 
\begin{equation}
\begin{aligned}
    f=&\frac{1}{2\pi\sqrt{\det\Sigma}}
\exp\!\left[-\frac{1}{2}\left(v_\bmk,\frac{v^{\prime}_\bmk}{\cs k}\right)\,\Sigma^{-1}\left(v_\bmk,\frac{v^{\prime}_\bmk}{\cs k}\right)^{\!\top}\right].
\end{aligned}
\end{equation}
Introducing polar coordinates $v_\bmk\equiv \rho_\bmk\cos\theta_\bmk$ and $v^{\prime}_\bmk/(\cs k)\equiv \rho_\bmk\sin\theta_\bmk$, with Jacobian $J=\rho_\bmk$, and marginalising over $\rho_\bmk$, the probability density for $\theta_\bmk$, in terms of the squeezing parameter, takes the form
\begin{equation}\label{eq:pdf_theta}
f(\theta_\bmk)=\frac{1}{\pi\left(\cosh 2r_k-\sinh 2r_k\,\cos 2\theta_\bmk\right)}\,.
\end{equation}
The distribution function is even in $\theta_\bmk$, so its mean vanishes and its dispersion reads~\cite{Vareilles:2026acg}
\begin{equation}\label{eq:Delta_theta_superHmain}
    (\Delta\theta_k)^2  = \frac{\pi^2}{12} + \text{Li}_2\left(- \tanh r_k\right) \underset{r_k \gg 1}{\simeq} 2 \log(2) \ee^{-2r_k}\,.
\end{equation}

In the super-horizon limit, the covariance is dominated by the momentum spectrum, $\ee^{2r_k}\simeq2|v_k'|^2/(\cs k)$, and the dispersion~\eqref{eq:Delta_theta_superHmain} takes the form
\begin{equation}\label{eq:dtheta_modefunctions}
(\Delta\theta_k)^2\simeq\log(2)\;\frac{\cs k}{|v_k'|^{2}}\,.
\end{equation}
At leading order in slow roll, the squeezing amplitude is twice the number of $e$-folds $N_{\mathrm{inf}}(k)$ that the mode $k$ spends outside the Hubble radius during inflation, $r_k\simeq 2N_{\mathrm{inf}}(k)$, so $\Delta\theta_k\sim\ee^{-2N_{\mathrm{inf}}(k)}$.
Inflation thus produces highly squeezed states, and
eq.~\eqref{eq:Delta_theta_superHmain} makes explicit the intuition that more squeezing implies better coherence~\cite{Albrecht:1996dq}, with coherence sharpening exponentially in $N_{\rm inf}(k)$. Scenarios at different energy scales, say Grand Unified Theory (GUT) versus electroweak (EW) inflation, differ by nearly a factor of two in their $e$-fold count $N_{\rm inf}(k)$, and would therefore be expected to imprint drastically different phase coherences.\\


\paragraph*{Universality of phase coherence.---}

A common expectation is that phase coherence directly reflects the cumulative squeezing acquired during inflation, and therefore retains a memory of the inflationary energy scale through the number of $e$-folds a comoving mode spends outside the Hubble radius. We show that this intuition is incomplete: once the post-inflationary evolution is consistently included, the correspondence between longer super-Hubble evolution and sharper coherence no longer holds.

We model transitions between cosmological eras as the crossing of instantaneous, constant-density hypersurfaces crossings in an FLRW spacetime, enforcing continuity of the induced metric and extrinsic curvature on a constant-density hypersurface. At the background level this gives $[a]_\pm=0$ and $[\mathcal{H}]_\pm=0$ where ${\cal H}\equiv a'/a$. At the perturbative level, we show in the Supplementary Material that the Deruelle--Mukhanov conditions \cite{Deruelle:1995kd} translate into junction conditions for the Mukhanov--Sasaki variable,
\begin{equation}
\label{eq:matching}
\begin{aligned}
\left[\frac{1}{z\,\cs ^2}\left(v_k^{\prime}  - \frac{z'}{z}v_k - 3\mathcal{H}\cs ^2 v_k\right)\right]_\pm &= 0\,,\\
\left[\frac{1+w}{z\,\cs ^2}\left(v_k^{\prime}  - \frac{z'}{z}v_k\right)\right]_\pm &= 0\,,
\end{aligned}
\end{equation}
which define a linear map $\boldsymbol{V}_k^+ = \boldsymbol{M}\boldsymbol{V}_k^-$ on the canonical vector $\boldsymbol{V}_k=(v_k,v_k^{\prime})^{\top}$, between the left ($-$) and right ($+$) side of the transition hypersurface. The matrix $\boldsymbol{M}$ is symplectic, so each transition acts effectively as a Bogoliubov transformation mixing growing and decaying modes and reshuffling the squeezing parameters.

During inflation, we assume a quasi-de Sitter phase with $a(\eta) \propto \eta^{-1}$, together with Bunch--Davies initial conditions.  During radiation domination, $a(\eta)\propto\eta$, so that $z''/z=0$: the oscillator frequency is constant, no Bogoliubov mixing occurs, and $r_k$ freezes exactly. During matter domination, we retain the residual sound speed of the matter--radiation mixture at leading order in $a_{\rm eq}/a$. This finite $\cs ^2$ preserves acoustic sub-Hubble dynamics, whereas setting $\cs ^2=0$ would render the Mukhanov--Sasaki description singular. Details are provided in the Supplemental Material.

Within each era, $-\log\Delta\theta_k$ grows linearly on super-Hubble scales with the number of $e$-folds, with a rate that depends on the background equation of state. This is precisely what is observed in Fig.~\ref{fig:mainfig}, which nonetheless reveals a much stronger statement: two scenarios differing only by their inflationary scale $H_{\rm inf}$, and therefore by both the duration of inflation and the value of the matching parameters entering the transfer matrix $\boldsymbol{M}$, yield \textit{identical} phase coherences at equality and thereafter at the CMB. The curves in fact already coincide during radiation domination. In particular, all explicit dependence on $H_{\rm inf}$ that enters through the inflationary squeezing parameters, reheating matching conditions, and the subsequent radiation-dominated evolution cancels out in the final result.\\


\paragraph*{Origin of the universality.---}

\begin{figure}[t]
    \centering
        \centering
        \includegraphics[width=1.\linewidth]{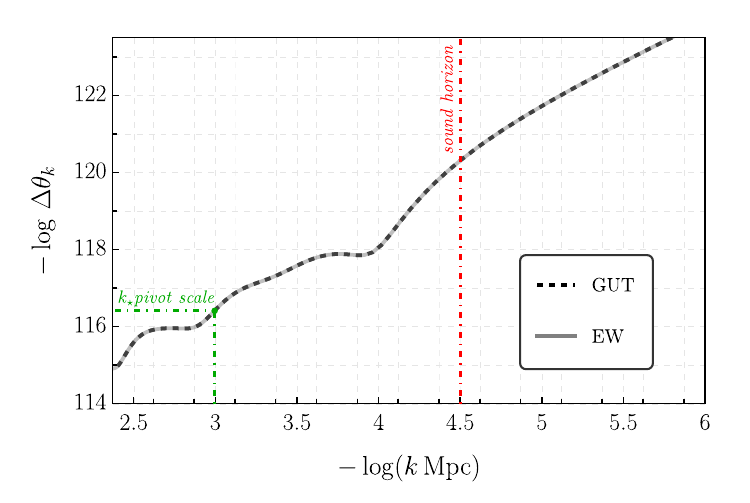}
    \caption{Logarithmic phase variance $-\log\Delta\theta_k$ at recombination ($\eta=\eta_{\rm CMB}$) for EW (solid) and GUT (dashed) inflation. The red line marks the sound horizon, separating super-horizon (right) from sub-horizon (left) modes; the green line marks the pivot scale $k_\star=0.05\,\mathrm{Mpc}^{-1}$. $\Delta\theta_k$ improves with scale above the horizon, reflecting the accumulated $e$-folds, and oscillates below the sound horizon, where it sets the phase coherence of the CMB. The two scenarios coincide at all $k$: $-\log\Delta\theta_k$ is independent of the inflationary energy scale, with $-\log\Delta\theta_{k_\star}\simeq117$ at the pivot scale.}
    \label{fig:overall2}
\end{figure}

The phase coherence at the end of the radiation era collects three contributions: the coherence built up gradually during inflation, the discrete jump imprinted at reheating and the subsequent evolution through the radiation era. The universality lies in how these three pieces scale with the inflationary energy scale. For a mode spending $N_{\rm inf}(k)$ $e$-folds outside the Hubble radius, the phase spread sharpens as $-\log\Delta\theta_k\simeq2 N_{\rm inf}$ at leading order in slow roll. Let us first assume that reheating is instantaneous and that the radiation era starts right at the end of inflation (this assumption will be relaxed below where our result will be shown to generalise to any reheating kinematics~\cite{Martin:2006rs, Martin:2010kz, 2012PhRvD..85j3533E, Martin:2014nya, Martin:2016oyk}).
The matching conditions thus imply that $\Delta\theta_k\sqrt{(1+w)/\cs^3}$ is constant at the transition (see Supplementary Material).
The induced contribution to $\Delta\theta_k$ hence scales as $\sqrt{1+w_{\mathrm{inf}}}$. Using $1+w_{\mathrm{inf}}=2\epsilon_1/3$, where $\epsilon_1\equiv1-\mathcal{H}'/\mathcal{H}^2$ is the first slow-roll parameter, this implies that $\Delta\theta_k$ receives a contribution $\propto\sqrt{\epsilon_1}$ from instantaneous reheating.
The smaller $\epsilon_1$, the closer $w$ lies to $-1$ before the transition, and the stronger the phase coherence. At last, during radiation the spread evolves as $\Delta\theta_k\propto a^{(1-3w_{\rm rad})/2}$ (for pure radiation, $w_{\rm rad}=1/3$, it freezes). Hence, combining the contributions from inflation, reheating and the radiation dominated era, we obtain at the radiation--matter equality
\begin{equation}\label{eq:logdtheta_scaling}
    -\log\Delta\theta_k\big\rvert_{\mathrm{eq}}
    \simeq 2 N_{\rm inf}-\tfrac12\log\epsilon_1+\tfrac{1}{2} (1-3w_{\rm rad})N_{\rm rad}+\text{const}\,.
\end{equation}

Each of these terms depends on the inflationary energy scale, yet the total sum does not.
Since $H\propto a^{-(3+3w_{\rm rad})/2}$ throughout radiation, $Ha^{(3+3w_{\rm rad})/2}$ is conserved and fixed by the observed radiation density today. It follows that $a_{\rm rh}\sim H_{\rm rh}^{-2/(3+3w_{\rm rad})}$: a higher reheating scale corresponds to a smaller scale factor, hence a larger redshift at reheating. The radiation era, extending from $a_{\rm rh}$ to the fixed matter–radiation equality scale $a_{\rm eq}$, therefore lasts
\begin{equation}
N_{\rm rad}
=\log\left(\frac{a_{\rm eq}}{a_{\rm rh}}\right)
\simeq\frac{2}{3+3w_{\rm rad}}
\log H_{\rm inf}+\text{const}\,,
\end{equation}
under the assumption of instantaneous reheating, with $H_{\rm inf}=H_{\rm rh}$. 
The same scaling sets the inflationary duration:
a fixed comoving mode exits the horizon at $k=a_{\rm exit}H_{\rm inf}$ during inflation, so
$a_{\rm exit}\sim H_{\rm inf}^{-1}$, such that
\begin{equation}
N_{\rm inf}
=\log\left(\frac{a_{\rm rh}}{a_{\rm exit}}\right)
\simeq\frac{1+3w_{\rm rad}}{3+3w_{\rm rad}}
\log H_{\rm inf}+\text{const}\,.
\end{equation}
Combined with Eq.~\eqref{eq:logdtheta_scaling}, the coefficients of $\log H_{\rm inf}$ sum up to $(2+6w_{\rm rad})/(3+3w_{\rm rad}) + (1-3w_{\rm
rad})/(3+3w_{\rm rad}) = 1,$
and the inflationary scale therefore enters only as
\begin{equation}\label{eq:logdtheta_cancel}
    -\log\Delta\theta_k\big\rvert_{\rm eq}\simeq \tfrac12\log\left(\frac{H_{\rm inf}^2}{\epsilon_1}\right)+\text{const}\,.
\end{equation}
This combination is precisely fixed by the measured amplitude of the primordial power spectrum, $\AS \propto H_{\rm inf}^2/\epsilon_1$ evaluated at horizon crossing.
Thus, a higher energy scale is bound to a larger $\epsilon_1$ to keep $\AS $ fixed; the two variations compensate, and only $\AS$ remains. This binding is the core mechanism behind the independence from the inflationary energy scale. For $w_{\rm rad}=1/3$ the radiation contribution in~\eqref{eq:logdtheta_scaling} vanishes, so this universality already holds at reheating, as in Fig.~\ref{fig:mainfig}. A complete computation is performed in the Supplemental Material that retains all prefactors and slow-roll corrections, and confirms the cancellation. \\


\paragraph*{Generality from adiabatic evolution.---}

Remarkably, the equation of state $w_{\rm rad}$ has also cancelled out of the final result~\eqref{eq:logdtheta_cancel} of the previous section: although it controls both the duration of the post-inflationary era and the growth rate of the phase uncertainty during it, the dependences on the inflationary energy scale compensate exactly for any value of $w_{\rm rad}$, and the coherence evaluated at a late reference time is insensitive to it. We now show that this is not an accident, but a general property of adiabatic evolution. 

On superhorizon scales, the phase uncertainty satisfies eq.~\eqref{eq:dtheta_modefunctions}, while conservation of the curvature perturbation implies $|v_k'|= \mathcal{H}\,z\,|\zeta_k|$, up to corrections of order $ w^{\prime}/[\mathcal{H}(1+w)]$ and $ \cs^{\prime}/(\mathcal{H}\cs)$ which are negligible whenever the equation of state and sound speed vary slowly on a Hubble time. 
Expressing $|\zeta_k|^2$ in terms of the primordial power spectrum,
\begin{align}
|\zeta_k|^2=
\frac{2\pi^2}{k^3}\,
 \AS
\left(\frac{k}{k_\star}\right)^{n_s-1},
\end{align}
immediately yields
\begin{align}\label{eq:logdtheta_general}
(\Delta\theta_k)^2
=
\frac{\log (2)\,\cs^3}{6\pi^2(1+w)\,\AS (k/k_\star)^{n_s-1}}
\,\frac{H^2}{M_{\rm Pl}^2}
\left(\frac{k}{aH}\right)^4\,.
\end{align}

Eq.~\eqref{eq:logdtheta_general} is the central result of this Letter. Two assumptions enter: the curvature perturbation is conserved, as for
adiabatic super-Hubble perturbations whose non-constant mode decays~\cite{Wands:2000dp,Weinberg:2003sw,Lyth:2004gb}; and the mode
evolves unitarily. Beyond the amplitude of the primordial power spectrum, no trace of the inflationary energy scale, of the duration of inflation, or of the intermediate expansion history survives in phase coherence: evaluated at any observationally specified super-Hubble instant, the background quantities are themselves fixed by the observed late-time cosmology, so that scenarios differing by their inflationary scale have converged and
remain identical thereafter.
This is why the above assumption of instantaneous reheating was in fact not necessary. 
As a consequence, arbitrarily exotic expansion histories, such as prolonged preheating, early matter- or primordial-black-hole domination, kination, episodes of early dark energy, or any intermediate era with time-dependent equation of state, leave $(\Delta\theta_k)^2$ unchanged. 

Eq.~\eqref{eq:logdtheta_general} hence provides a consistency relation for single-field inflation with conserved $\zeta$, linking the phase uncertainty of primordial perturbations to the amplitude of their power spectrum. 
A significant loss of coherence relative to Eq.~\eqref{eq:logdtheta_general} would indicate new primordial physics, such as entropy perturbations breaking adiabaticity or decoherence breaking unitarity. \\

\paragraph{Phase Coherence in the CMB.---}

Having established the universality of $\Delta \theta_k$ on superhorizon scales, we now follow the coherence itself throughout horizon re-entry until recombination, where the CMB anisotropies are imprinted. Figure~\ref{fig:overall2} shows $\Delta \theta_k$ across comoving scales at fixed $\eta_{\rm CMB}$. On super-horizon scales it decays monotonically as $k$ decreases, since a larger-scale mode spends more $e$-folds outside the horizon and is more strongly squeezed~\cite{Dodelson:2003ip}. On scales below the sound horizon at recombination, $\Delta\theta_k$ measures the dispersion of the acoustic oscillation phase, and oscillates with $k$.

\begin{figure}[t]
    \centering
    \includegraphics[width=1.\linewidth]{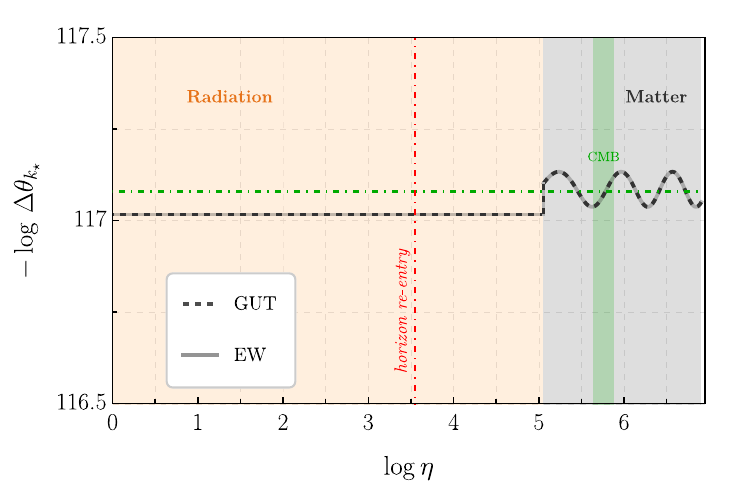}
    \caption{Time evolution of the logarithmic phase variance $-\log\Delta\theta_{k_{\star}}$ at the pivot scale $k_\star$, across the radiation-dominated and matter-dominated eras, for EW (\textit{solid}) and GUT (\textit{dashed}) inflationary scenarios. The vertical red dashed line indicates sound horizon crossing, while the green shaded region corresponds to the thickness of the last scattering surface translated in conformal time, centred on recombination $\eta_{\rm CMB}$, yielding $-\log\Delta\theta_{k_*} \simeq 117$.}
    \label{fig:placeholder}
\end{figure}

At the pivot scale, the electroweak and GUT scenarios both give $-\log\Delta\theta_{k_\star}\simeq117$, confirming universality at CMB scales. Figure~\ref{fig:placeholder} follows its evolution: the coherence is constant throughout the radiation era and oscillates about a constant value in the matter era, so that horizon re-entry does not degrade $\Delta\theta_k$ established outside the horizon (see also Supplementary Material). The last-scattering surface has a finite width $\Delta z_*/z_* \sim 90/1090 \sim 8.25\%$ (green band) over which the phase coherence is averaged. We stress that these are not oscillations of the phase itself, but rather of its variance. One should indeed distinguish the acoustic oscillation, which corresponds to oscillations in the deterministic evolution of the \emph{mean} phase $\phi_k$ (see footnote~\ref{footnote:remove:phik}), from the oscillations in Figure~\ref{fig:placeholder} that characterise its \emph{variance} and which have no classical counterpart.

In summary, the CMB phase coherence at the pivot scale, $-\log\Delta\theta_{k_\star}\simeq117$, is a universal prediction of slow-roll inflation: fixed by the primordial power spectrum alone, it persists through the radiation era and horizon re-entry to recombination. \\

\paragraph*{Outlook.---} 

In this Letter, we have shown that the phase coherence of cosmological perturbations at recombination is not solely determined by inflationary squeezing, but arises from a non-trivial interplay between inflationary evolution and subsequent cosmological dynamics. Once the full cosmological history is taken into account, the dependence on the inflationary energy scale and on the post-inflationary evolution (including reheating) cancels, yielding a universal prediction fixed entirely by the observed amplitude of the primordial power spectrum. This result revises the standard expectation that higher-energy inflation generically produces stronger coherence, and emphasizes the crucial role of post-inflationary evolution in shaping observable CMB features.

The amount of coherence of the acoustic peaks is thus a generic, universal prediction of adiabatic and unitary evolution during inflation, rather than a model-dependent feature.
Its relation to the amplitude of the temperature anisotropies, see eq.~(\ref{eq:logdtheta_general}), constitutes a new ``consistency relation'' of inflation (similar in spirit to the one that exists between the amplitude and the tilt of the tensor power spectrum for instance~\cite{Lidsey:1995np}).
Deviations from this pattern, such as enhanced mode incoherence, peak broadening, or distortions of the $TE$ phase relation, would provide direct evidence for physics beyond this minimal framework. Environmental interactions, for instance, can induce decoherence, reducing the suppression of the decaying mode and thereby degrading phase coherence~\cite{Martin:2021znx,Martin:2022kph,Colas:2024ysu,Colas:2024xjy,Micheli:2025yux,Haque:2026bby}. Multi-field, warm, dissipative, and gauge inflation provide concrete settings in which this may occur, through entropy transfer and/or the dissipation and noise generated by additional degrees of freedom~\cite{Berera:1995ie,Berera:2008ar,Berghaus:2025dqi,Anber:2009ua,Peloso:2022ovc,vonEckardstein:2023gwk,Creminelli:2023aly,Salcedo:2024smn, Colas:2025ind,Cespedes:2026fdp,Salcedo:2026sdn}. A complementary possibility is that of non-attractor evolution that revives the nominally decaying mode, as during transient departures from slow roll e.g. in ultra-slow-roll phases~\cite{Inoue:2001zt,Kinney:2005vj,dePutter:2019xxv}. \\

\paragraph*{Acknowledgments.---} 
We thank William Coulton, Amaury Micheli, Oliver Philcox, Xi Tong and Yong Sheng Yap for insightful discussions. T.C. acknowledges the Julian Schwinger Foundation for financial support to attend the 2026 Peyresq Spacetime Meeting, where valuable discussions contributed to aspects of this work. This work has been supported by STFC consolidated grants ST/X001113/1, ST/T000694/1, ST/X000664/1, ST/Y509127/1 and EP/V048422/1. F.P. and S.V. received support from the French government under the France 2030 investment plan, as part of the Initiative d'Excellence d'Aix-Marseille Universit\'e - A*MIDEX (AMX-19-IET-012) and by the ``action th\'ematique" Cosmology-Galaxies (ATCG) of the CNRS/INSU PN Astro and by the {\it Agence Nationale de la Recherche} under the grant ANR-24-CE31-6963-01.

\bibliographystyle{apsrev4-2}
\bibliography{biblio}

\clearpage

\appendix

\begin{widetext}

\section{SUPPLEMENTARY MATERIAL
}

This Supplementary Material provides technical details underlying the results of the main text. We first present the relation between the Mukhanov-Sasaki variable and the phase coherence through a WKB approximation. We then compute the moments of the phase coherence $\theta_{\bmk}$ and their expansion in the large-squeezing limit. We provide details on the Mukhanov-Deruelle matching conditions and the modelling of the different cosmological eras. At last, explicitly compute the jumps experienced by the various quantities of interest at the reheating surface and exhibit the independence of the phase coherence from the inflationary energy scale.


\section{Derivation of the WKB expansion}\label{WKB}

\paragraph{WKB ansatz.}

Assuming $v_k(\eta) = A_k(\eta) \exp[i B_k(\eta)]$ with real amplitude and phase, the imaginary part of \eqref{eq:fourier_modes} enforces $2A_k' B_k' + A_k B_k'' = 0$, which can be rewritten as $(A_k^2 B_k')' = 0$. This requires $A_k \propto 1/\sqrt{|B_k'|}$; defining $W_k(\eta) \equiv |B_k'(\eta)|$ and using the canonical Wronskian condition $v_k v_k^{* \prime} - v_k^* v_k^\prime = i$, the WKB ansatz takes the form
\begin{equation}
v_k^{\pm}(\eta) = \frac{1}{\sqrt{2\,W_k(\eta)}}\,
\exp\!\left[\,\mp\,i\!\int^{\eta}\! W_k(\eta')\,\dd\eta'\,\right],
\label{eq:WKBansatz}
\end{equation}
where the two signs label the two independent solutions.
Substituting \eqref{eq:WKBansatz} back into the real part of \eqref{eq:fourier_modes} yields the equation
\begin{equation}
W_k^{\,2} \;=\; \omega_k^2 \;+\; \frac{3\,(W_k')^2}{4\,W_k^{\,2}} \;-\; \frac{W_k''}{2\,W_k},
\label{eq:Wexact}
\end{equation}
which we solve iteratively in the adiabaticity parameter
$\epsilon_{\rm WKB} \equiv |\omega_k'/\omega_k^{\,2}| \ll 1$
(which enforces that the amplitude evolves much more slowly than the phase, $|A_k'/A_k| \ll |B_k'|$).

\paragraph{Leading order.} Dropping the RHS corrections $W_k^{\prime}$ in \eqref{eq:Wexact} sets $W_k^{(0)} = \omega_k$, giving
\begin{equation}
v_k^{(0),\pm}(\eta) \;=\; \frac{1}{\sqrt{2\,\omega_k(\eta)}}\,
\exp\!\left[\,\mp\,i\!\int^{\eta}\!\omega_k(\eta')\,\dd\eta'\,\right].
\label{eq:LO}
\end{equation}

\paragraph{Next-to-leading order.} Inserting $W_k = \omega_k$ on the RHS of \eqref{eq:Wexact} and taking the square root to first order in $\epsilon_{\rm WKB}^{\,2}$,
\begin{equation}
W_k(\eta) \;=\; \omega_k(\eta) \;+\; \delta\omega_k(\eta), \qquad
\delta\omega_k(\eta) \;\equiv\; \frac{3\,\left[\omega_k'(\eta)\right]^{2}}{8\,\omega_k^{\,3}(\eta)} \;-\; \frac{\omega_k''(\eta)}{4\,\omega_k^{\,2}(\eta)}.
\label{eq:deltaomega}
\end{equation}

\paragraph{Final approximate solution.}
\begin{equation}
\;
v_k^{\pm}(\eta) \;=\; \frac{1}{\sqrt{2\,\omega_k(\eta)}}
\left[\,1 \;-\; \frac{\delta\omega_k(\eta)}{2\,\omega_k(\eta)}\,\right]
\exp\!\left\{\,\mp\,i\!\int^{\eta}\!\left[\omega_k(\eta') + \delta\omega_k(\eta')\right]\,\dd \eta'\,\right\}\;
\label{eq:vk_final}
\end{equation}
The two independent modes correspond to the $\mp$ choice in the exponent; their real linear combination,
\begin{equation}
v_k(\eta) \;=\; \frac{1}{\sqrt{\omega_k(\eta)}}\!\left(1-\frac{\delta\omega_k}{2\,\omega_k}\right)\!
\left[\,\alpha_k\cos\Phi_k(\eta) + \beta_k\sin\Phi_k(\eta)\,\right],\qquad
\Phi_k(\eta) \equiv \int^{\eta}\!\left(\omega_k+\delta\omega_k\right)\,\dd\eta',
\end{equation}
makes the oscillatory structure explicit.

 The combination $\alpha_k\cos\Phi_k + \beta_k\sin\Phi_k$ in \eqref{eq:vk_final} can be re-expressed as a single cosine with offset phase via the identity $\alpha\cos\Phi + \beta\sin\Phi = \mathcal{A}\cos(\Phi - \theta)$, where $\mathcal{A}=\sqrt{\alpha^2+\beta^2}$ and $\tan\theta = \beta/\alpha$. The final WKB solution then takes the compact form
\begin{equation}
\;
v_k(\eta) \;=\; \frac{\mathcal{A}_k}{\sqrt{\omega_k(\eta)}}
\left[\,1 \;-\; \frac{\delta\omega_k(\eta)}{2\,\omega_k(\eta)}\,\right]
\cos\left[\,\Phi_k(\eta) \;-\; \theta_k\,\right]\;
\label{eq:vk_phase}
\end{equation}
with the $k$-dependent constants
\begin{equation}
\mathcal{A}_k \;\equiv\; \sqrt{\alpha_k^{\,2}+\beta_k^{\,2}},
\qquad
\tan\theta_k \;\equiv\; \frac{\beta_k}{\alpha_k}\,.
\end{equation}
The two integration constants $(\alpha_k,\beta_k)$ are repackaged as an \emph{amplitude} $\mathcal{A}_k$ and a \emph{phase offset} $\theta_k$.

To extract the phase explicitly, we write~\eqref{eq:vk_phase} as
\begin{equation}
v_k(\eta)=\mathcal B_k(\eta)\cos\left[\Phi_k(\eta)-\theta_k\right],
\qquad
\mathcal B_k(\eta)\equiv \frac{\mathcal A_k}{\sqrt{\omega_k(\eta)}}
\left[1-\frac{\delta\omega_k(\eta)}{2\omega_k(\eta)}\right].
\end{equation}
A direct differentiation gives
\begin{equation}\label{eq:theta_arctan_NLO}
\frac{v_k'}{\omega_k v_k}
=
\frac{\mathcal B_k'}{\omega_k \mathcal B_k}
-\frac{\Phi_k'}{\omega_k}\tan\left(\Phi_k-\theta_k\right),
\end{equation}
so that, using $\Phi_k'=\omega_k+\delta\omega_k$ and
$\mathcal B_k'/\mathcal B_k= -\omega_k'/(2\omega_k)+\mathcal O(\epsilon_{\rm WKB}^2\omega_k)$, one obtains
\begin{equation}
-\frac{v_k'}{\omega_k v_k}
=
\tan\left[\Phi_k(\eta)-\theta_k\right]
+\mathcal O(\epsilon_{\rm WKB})\,.
\end{equation}
Therefore the extra term $\omega_k'/(2\omega_k^2)$ appearing in~\eqref{eq:theta_arctan_NLO} is adiabatically suppressed, since it is of order $\epsilon_{\rm WKB}\equiv |\omega_k'/\omega_k^2|\ll1$, whereas the leading oscillatory contribution to $-v_k'/(\omega_k v_k)$ is order one. As a consequence, the phase can be written at leading order as
\begin{equation}
\theta_k(\eta)=\arctan\!\left(-\frac{v_k'}{\omega_k v_k}\right)
+\mathcal O(\epsilon_{\rm WKB})
\pmod{\pi}\,.
\end{equation}


\section{Moments of \texorpdfstring{$\theta_{\bmk}$}{θk} and the large-squeezing limit}

Using $\cosh2r_k-\sinh2r_k\cos2\theta_\bmk
=\ee^{-2r_k}\cos^2\theta_\bmk+\ee^{2r_k}\sin^2\theta_\bmk$, the
density~\eqref{eq:pdf_theta} takes exactly the form of the phase distribution
of a pure Gaussian pair~\cite{Vareilles:2026acg},
\begin{equation}\label{eq:pdf_match}
f(\theta_\bmk)=\frac{\cs k}{2\pi}\left[
\widetilde P_{pp}\cos^2\theta_\bmk
-2\cs k\widetilde P_{vp}\sin\theta_\bmk\cos\theta_\bmk
+\cs^2k^2\widetilde P_{vv}\sin^2\theta_\bmk\right]^{-1},
\end{equation}
with spectra
\begin{equation}\label{eq:tilde_spectra}
\cs k\,\widetilde P_{vv}=\frac{\ee^{2r_k}}{2}\,,\qquad
\frac{\widetilde P_{pp}}{\cs k}=\frac{\ee^{-2r_k}}{2}\,,\qquad
\widetilde P_{vp}=0\,,
\end{equation}
which satisfy the purity condition
$\widetilde P_{vv}\widetilde P_{pp}-\widetilde P_{vp}^2=1/4$. The closed-form
moments of~\cite{Vareilles:2026acg} then apply directly. The mean,
$\arctan\!\big[2\widetilde P_{vp}/(2\cs k\widetilde P_{vv}+1)\big]$, vanishes
identically since $\widetilde P_{vp}=0$, consistently with the parity
of~\eqref{eq:pdf_theta}, while the argument of the dilogarithm in the
variance is real,
\begin{equation}\label{eq:var_derivation}
(\Delta\theta_k)^2=\frac{\pi^2}{12}
+{\rm Li}_2\!\left(1-\frac{4\cs k\widetilde P_{vv}}
{1+2\cs k\widetilde P_{vv}}\right)
=\frac{\pi^2}{12}
+{\rm Li}_2\!\left(\frac{1-\ee^{2r_k}}{1+\ee^{2r_k}}\right)
=\frac{\pi^2}{12}+{\rm Li}_2\!\left(-\tanh r_k\right)\,,
\end{equation}
which is Eq.~\eqref{eq:Delta_theta_superHmain} of the main text. At large
squeezing, $\tanh r_k=1-2\ee^{-2r_k}+\mathcal O(\ee^{-4r_k})$ and, expanding
about ${\rm Li}_2(-1)=-\pi^2/12$ with
$\dd\,{\rm Li}_2(-x)/\dd x=-\log(1+x)/x$,
\begin{equation}
(\Delta\theta_k)^2=2\log2\,\ee^{-2r_k}-\ee^{-4r_k}
+\mathcal O\!\left(\ee^{-6r_k}\right)\,.
\end{equation}


\section{Matching conditions}\label{subsec:matching}

To model instantaneous transitions between cosmological eras, such as from inflation to radiation domination, or from radiation to matter domination, we use the \emph{junction conditions}~\cite{MSM_1927__25__1_0, Israel:1966rt} between two FLRW spacetimes, which require the continuity of the induced three-metric and the extrinsic curvature across a transition hypersurface $\Sigma$. In our case, on each side of $\Sigma$, the universe is described by a perfect fluid with equation of state parameter $w$ and sound speed $\cs $, with subscripts $-$ and $+$ denoting quantities evaluated immediately before and after the transition, respectively. The transition hypersurface is taken to be one of constant energy density $\rho=\rho_0+\delta\rho=\text{const}$. 

At the background level, continuity of the induced metric and extrinsic curvature implies
\begin{equation}
    \left[a\right]_{\pm}=0,\qquad\left[\mathcal{H}\right]_{\pm}=0\,,
\end{equation}   
where ${\cal H}\equiv a'/a$ and $[X]_\pm \equiv X_+ - X_-$ denotes the jump of any quantity $X$ across $\Sigma$.
At the level of perturbations, Deruelle and Mukhanov~\cite{Deruelle:1995kd} showed that, in Fourier space, they translate into matching conditions for the gauge-invariant perturbations
\begin{align}\label{matching_condition_gauge_invariant}
    \left[\zeta_k+\frac{2}{9(1+w)}\frac{k^2}{\mathcal{H}^2}\Phi_k\right]_{\pm}=0\,,
    \qquad [\Phi_k]_{\pm}=0\,,
\end{align}
where $\Phi$ is the Bardeen potential (using the Friedmann relation $\mathcal{H}'-\mathcal{H}^2=-\tfrac32(1+w)\mathcal{H}^2$ and, in Fourier space, $\Delta\to-k^2$) and $\zeta$ is the comoving curvature perturbation, $\zeta\equiv\Phi+\frac{\mathcal{H}}{\bar\phi'}\,\delta\phi$, i.e. the comoving curvature perturbation (often denoted $\mathcal{R}$; we follow the convention of Maldacena~\cite{Maldacena:2002vr}, in which $\zeta$ is the curvature perturbation on comoving slices).  We assume no anisotropic stress, so the two metric potentials coincide, $\Phi=\Psi$. The bracketed quantity is precisely the curvature perturbation on uniform-density slices, $\zeta^{\rm ud}_k\equiv\zeta_k+\frac{2}{9(1+w)}\frac{k^2}{\mathcal{H}^2}\Phi_k$ (the Bardeen--Steinhardt--Turner variable~\cite{Bardeen:1983qw}, which coincides with $\zeta$ in the super-horizon limit), so~\eqref{matching_condition_gauge_invariant} enforces continuity of $\Phi$ and of $\zeta^{\rm ud}$ across the transition; $\zeta$ itself is then continuous only up to $\mathcal{O}(k^2/\mathcal{H}^2)$, the residual jump being set by the discontinuity in $w$. We now derive the equivalent junction conditions for the Mukhanov--Sasaki variable from~\eqref{matching_condition_gauge_invariant}.

The comoving curvature perturbation $\zeta_k$ is related to the Bardeen potential $\Phi_k$ by
\begin{align}
    \zeta_k &= \Phi_k + \frac{2}{3(1+w)}\left(\mathcal{H}^{-1}\Phi_k' + \Phi_k\right),\label{eq:app_R1}\\
    \mathcal{H}^{-1}\zeta_k' &= -\frac{2}{3(1+w)}\left(\frac{\cs  k}{\mathcal{H}}\right)^2\Phi_k\,,\label{eq:app_R2}
\end{align}
where in~\eqref{eq:app_R2} we used the adiabaticity of the fluid, $\cs ^2=w$, which removes any non-gradient source.
These can be combined with the matching conditions~\eqref{matching_condition_gauge_invariant} to obtain junction conditions directly for $\zeta_k$ and its derivative. Substituting~\eqref{eq:app_R2} into the first of~\eqref{matching_condition_gauge_invariant} to eliminate $\Phi_k$ in favour of $\zeta_k'$ gives
\begin{equation}\label{eq:app_matching_zeta_1}
    \left[\zeta_k - \frac{1}{3\mathcal{H}\cs ^2}\zeta_k'\right]_{\pm} = 0\,,
\end{equation}
while the continuity of $\Phi_k$ together with~\eqref{eq:app_R2} and the background matching condition $[\mathcal{H}]_\pm = 0$ yields
\begin{equation}\label{eq:app_matching_zeta_2}
    \left[\frac{1+w}{\cs ^2}\zeta_k'\right]_{\pm} = 0\,.
\end{equation}

To translate these into conditions on the Mukhanov--Sasaki variable, we use the relations
\begin{equation}
    \zeta_k = \frac{v_k}{z}\,, \qquad \zeta_k' = \frac{v_k' - \frac{z^{\prime}}{z}\,v_k}{z}\,,
\end{equation}
with $z^2 = 3M_{\mathrm{Pl}}^2(1+w)\,a^2/\cs ^2$. Substituting into~\eqref{eq:app_matching_zeta_1}--\eqref{eq:app_matching_zeta_2} and using $[a]_\pm = 0$, $[\mathcal{H}]_\pm = 0$, one obtains after algebraic simplification
\begin{align}&\left[\frac{1}{\cs ^2z}\left(v_k^{\prime}-\left(\frac{z^{\prime}}{z}+3{\cal H}\cs ^2\right)v_k\right)\right]_{\pm}=0\,, \label{matching_condition_v_1}\\
&\left[\frac{1+w}{\cs ^2z}\left(v_k^{\prime}-\frac{z^{\prime}}{z}v_k\right)\right]_{\pm}=0\,, \label{matching_condition_v_2}
\end{align}
which are the junction conditions used in the main text.
These two conditions define a linear canonical transformation $\boldsymbol{V}_k^+ = \boldsymbol{M}\,\boldsymbol{V}_k^-$ between the canonical pairs $\boldsymbol{V}_k^\pm = (v_k^\pm,\, v_k^{\prime\pm})^{\top}$ on either side of the transition. Since $\boldsymbol{M}$ is symplectic, it acts as a Bogoliubov transformation that generically modifies the squeezing parameters across $\Sigma$. Explicitly, the jump in squeezing amplitude $r_k$ is given by
\begin{equation}\label{eq:jump_squeezing}
    [r_k]_{\pm} = \frac{1}{2}\left\{\operatorname{arcosh}\!\left[\left(\boldsymbol{D}^+\boldsymbol{M}\boldsymbol{V}_k^-\right)^{\dagger}\left(\boldsymbol{D}^+\boldsymbol{M}\boldsymbol{V}_k^-\right)\right] - \operatorname{arcosh}\!\left[\left(\boldsymbol{D}^-\boldsymbol{V}_k^-\right)^{\dagger}\left(\boldsymbol{D}^-\boldsymbol{V}_k^-\right)\right]\right\}.
\end{equation}
where $\boldsymbol{D}^\pm = \mathrm{Diag}\!\left(\sqrt{\cs ^\pm\, k},\,\sqrt{(\cs ^\pm\, k)^{-1}}\right)$, which is a direct rewriting of the squeezing definition~\eqref{eq:squeezing} in matrix notation. The corresponding expression for the jump in the squeezing angle follows from the same logic applied to the expression of $\phi_k$ in~\eqref{eq:squeezing}
\begin{equation}\label{eq:jump_squeezing2}
    [\phi_k]_{\pm} = \frac{1}{2}\left[\operatorname{arctan}\frac{\left(\boldsymbol{M}\boldsymbol{V}_k^-\right)^{\dagger}\boldsymbol{J}\left(\boldsymbol{M}\boldsymbol{V}_k^-\right)}{\left(\boldsymbol{D}^+\boldsymbol{M}\boldsymbol{V}_k^-\right)^{\dagger}\boldsymbol{\eta}\left(\boldsymbol{D}^+\boldsymbol{M}\boldsymbol{V}_k^-\right)} - \operatorname{arctan}\frac{\left(\boldsymbol{V}_k^-\right)^{\dagger}\boldsymbol{J}\left(\boldsymbol{V}_k^-\right)}{\left(\boldsymbol{D}^-\boldsymbol{V}_k^-\right)^{\dagger}\boldsymbol{\eta}\left(\boldsymbol{D}^-\boldsymbol{V}_k^-\right)}\right]\,,
\end{equation}
where $\boldsymbol{J} = \begin{pmatrix} 0 & 1 \\ 1 & 0 \end{pmatrix}$ and 
$\boldsymbol{\eta} = \begin{pmatrix} 1 & 0 \\ 0 & -1 \end{pmatrix}$.


\section{Modelling the cosmological eras} 

In this work, we consider three distinct phases, i) inflation, ii) a radiation dominated era and iii) a matter dominated era for which we now derive the main quantities of interests.  

\paragraph{Inflation.} We consider a quasi-de Sitter background parameterized by the scale factor
\begin{equation}\label{eq:power_law_inflationS cale_factor}
    a(\eta)=l_0(-\eta)^{1+\beta}\,, \qquad  \eta \in ] - \infty, \, \eta_{\mathrm{rh}}]\,,
\end{equation}
with $\beta\lesssim-2$, where the limit $\beta=-2$ corresponds to exact de Sitter expansion. The Fourier-modes equation of motion~\eqref{eq:fourier_modes} reduce in that case to
\begin{equation}
    v_k''+\left[k^2-\beta(1+\beta)\eta^{-2}\right]v_k=0\,.
\end{equation}
Placing the modes in the Bunch-Davies vacuum in the deep-early inflationary era \cite{Bunch:1978yq}, the latter equation is solved for
\begin{equation}\label{eq:mode_inflation}
    v_k(\eta)=
\frac{\sqrt{\pi}}{2}(-\eta)^{1 / 2} H_\nu^{(1)}(-k \eta)
\end{equation}

In a strict power-law inflationary background, defined by the scale factor~\eqref{eq:power_law_inflationS cale_factor}, the first slow-roll parameter is constant and determined by $\epsilon_1 = \frac{2+\beta}{1+\beta}$. As a result, the second slow-roll parameter vanishes ($\epsilon_2 = 0$), restricting this exact solution to a small subset of inflationary potentials. However, in this work, we intentionally decouple the power-law index $\beta$ from the background slow-roll parameter $\epsilon_1$. We fix $\beta$ by matching to the observed scalar spectral index $\nS $, to effectively capture the non-zero $\epsilon_2$ dynamics of generic slow-roll models, while independently constraining $\epsilon_1$ via the observed scalar amplitude $\AS $. This phenomenological decoupling allows the resulting exact Hankel-like solutions of the Mukhanov-Sasaki equation to act as a generalized template for a wider class of slow-roll inflationary models than pure power-law inflation (moreover ruled out by the CMB) would conventionally permit.  


    \paragraph{Radiation dominated era.}

    
Imposing continuity of the scale factor and its derivative across the inflation-radiation transition at time $\eta_{\rm rh}$ yields the following profile
\begin{equation}\label{eq:scale_factor_RDE}
    a(\eta)=-l_0(-\eta_{\rm rh})^{\beta}(1+\beta)\left(\eta-\frac{\beta\eta_{\rm rh}}{1+\beta}\right)\,, \qquad  \eta \in ] \eta_{\mathrm{rh}}, \, \eta_{\rm eq}]\,,
\end{equation}
for the scale factor during the radiation-dominated era (RDE). The Fourier-mode functions equation reduces to 
\begin{equation}
    v_k''+\tfrac13k^2v_k=0\,,
\end{equation}
whose general solution is oscilatory and given by
\begin{equation}
    v_k(\eta)=C_{r1}\cos\left(\frac{k\eta}{\sqrt{3}}\right)+C_{r2}\sin\left(\frac{k\eta}{\sqrt{3}}\right)
\end{equation}
Taking as initial data the values inferred from the matching conditions~\eqref{matching_condition_v_1} and ~\eqref{matching_condition_v_2}, which fixes the constants $C_{r1}$ and $C_{r2}$, one can compute the power spectra and the squeezing parameters. 

\paragraph{Matter dominated era.} Imposing continuity of the scale factor and its derivative across the radiation-matter transition at time $\eta_{\rm eq}$ yields the following profile
\begin{equation}
a(\eta) = \frac{l_0 (-\eta_{\rm rh})^{\beta} \left(\eta + \beta \eta - 2 \beta \eta_{\rm rh} + \eta_{\rm eq} + \beta \eta_{\rm eq}\right)^2}{4 \beta \eta_{\rm rh} - 4 (1 + \beta) \eta_{\rm eq}}\,, \qquad  \eta \in ] \eta_{\rm eq}, \, \eta_0]\,,
\end{equation}
for the scale factor in the matter-dominated era (MDE), where $w(\eta)=\frac13\frac{a_{eq}}{a(\eta)}$ and $\cs ^2(\eta)=\frac{4}{9}\frac{a_{eq}}{a(\eta)}$ decrease with time.

Using a pressureless background with a non-vanishing sound speed is a consistent approximation if we interpret it as an expansion in $a_{\mathrm{eq}}/a \ll 1$. Indeed, in a matter--radiation mixture, the equation-of-state parameter and the sound speed are given by
\begin{equation}\label{eq:w_cs2_exact}
    w(\eta) = \frac{1}{3}\left[1+\frac{a(\eta)}{a_{\mathrm{eq}}}\right]^{-1}\,,\qquad
    \cs ^2(\eta) = \frac{1}{3}\left[1+\frac{3}{4}\,\frac{a(\eta)}{a_{\mathrm{eq}}}\right]^{-1}\,,
\end{equation}
both of which are $\mathcal{O}(a_{\mathrm{eq}}/a )$ and typically small $a_{\mathrm{eq}}/a \ll1$ in the deep matter-dominated era. The approximation scheme we are using treats these two quantities on an equal footing at $\mathcal{O}(a_{\mathrm{eq}}/a)$ and sets $w$ to \emph{exact zero} only when it appears in a combination where it is compared to a quantity of order unity. Concretely, in the Friedmann equation $2\mathcal{H}' = -\mathcal{H}^2(1+3w)$, the correction $3w$ is subleading relative to $1$, so setting $w=0$ yields the standard dust solution $a \propto \eta^2$ with relative error $\mathcal{O}(a_{\mathrm{eq}}/a)$. The same approximation applies to the combination $1/(1+w)=1+\mathcal{O}(a_{\mathrm{eq}}/a)$ appearing in the perturbation equations such as $z^2 \propto (1+w)\,a^2/\cs ^2$. Keeping $\cs ^2$ finite is physically necessary as it encodes the residual photon pressure that drives acoustic oscillations on sub-Hubble scales, and setting it to zero would artificially make the Mukhanov-Sasaki variable ill-defined. 

The solution to the mode functions~\eqref{eq:fourier_modes} are of Euler-type, since both terms of the effective frequency $\omega_k^2$ redshift identically, $\cs^2k^2\propto1/f^2$ and $z^{\prime\prime}/z\propto (f^{\prime})^2/f^2\propto1/f^2$. Hence, the solution is a pure power law of $f\propto\sqrt{a}$, which, in our setting, is explicitly given by 
\begin{equation}\label{solution_MS_MDE}
v_k(\eta)=  z_0 \sqrt{f(\eta)} \,
\Big[\,C_{m1}\,f(\eta)^{h/2}+C_{m2}\,f(\eta)^{-h/2}\Big]\,,
\end{equation}
with $z(\eta)= z_0 f^3$ and $f^2(\eta)=\mathcal{C}\,a(\eta)$ controlled by the (complex) numbers
\begin{equation}
\mathcal{C}
=
\frac{4\left[\beta\eta_{\rm rh}-(1+\beta)\eta_{\rm eq}\right]}
{l_0(-\eta_{\rm rh})^{\beta}}\,, \qquad
h = \frac{i^{-\beta} \eta_{\rm rh}^{-\beta/2} \sqrt{a_{\text{eq}}} k}{3 (1 + \beta)} \sqrt{\frac{\beta \eta_{\rm rh} - (1 + \beta) \eta_{\rm eq}}{l_0}} \mathcal{K}\,,
\end{equation}
and 
\begin{align}
    \mathcal{K} &= \sqrt{\frac{i^{2 \beta} (-\eta_{\rm rh})^{-\beta} \eta_{\rm rh}^{\beta} \left[225 l_0 (1 + \beta)^2 (-\eta_{\rm rh})^{\beta} + 64 a_{\text{eq}} k^2 \left((1 + \beta) \eta_{\rm eq} - \beta \eta_{\rm rh}\right)\right]}{a_{\text{eq}} k^2 (\beta \eta_{\rm rh} - (1 + \beta) \eta_{\rm eq})}}\,.
\end{align}
Using $\cs^2\propto1/a\propto1/f^2$ and $\mathcal H=2f'/f$, the ratio $\cs k/\mathcal H$ is constant throughout the matter era and may be evaluated once and for all at equality, $\cs k/\mathcal H=\tfrac23\,k/\mathcal H_{\rm eq}$. The exponent then takes the manifestly constant form
\begin{equation}\label{eq:h_const}
h^{2}=25-\frac{64}{9}\left(\frac{k}{\mathcal H_{\rm eq}}\right)^{\!2}
=25-16\left(\frac{\cs k}{\mathcal H}\right)^{\!2},
\end{equation}
real for $\cs k\ll\mathcal H$, giving growing and decaying solutions, and purely imaginary,
\begin{equation}\label{eq:nu_def}
h=2i\nu_k\,,\qquad
\nu_k=\sqrt{\frac{16}{9}\left(\frac{k}{\mathcal H_{\rm eq}}\right)^{\!2}
-\frac{25}{4}}\;,
\end{equation}
for sub-horizon $\cs k\gg\mathcal{H}$ modes, yielding oscillations in $\log f=\tfrac12\log a+{\rm const}$ at the fixed frequency $\nu_k$. Deep inside the sound horizon, one has $\nu_k\simeq\tfrac43\,k/\mathcal H_{\rm eq}$ and the instantaneous conformal frequency of $r_k$ is twice the acoustic frequency, $\nu_k\mathcal H\simeq2\cs k$. In this regime $|v_k|^2\propto f\,[\,{\rm const}+{\rm osc}\,]$, and the growing envelope $\propto f$ cancels against the redshifting weight $\cs k\propto1/f$ in the squeezing amplitude,
$\cosh2r_k=\cs k|v_k|^2+|v_k'|^2/(\cs k)={\rm const}+{\rm const}\times\cos\!\big(\nu_k\ln(a/a_{\rm eq})
+\varphi_{0,k}\big)$, with the two constants and the phase $\varphi_{0,k}$ fixed by $C_{m1}$ and $C_{m2}$: the squeezing oscillates with strictly constant amplitude about a constant mean.
Taking as initial data for the Fourier-modes the values inferred from the matching conditions~\eqref{matching_condition_v_1} and~\eqref{matching_condition_v_2} at the transition fixes the constants $C_{m1}$ and $C_{m2}$, we can compute the power spectra and the squeezing parameters in the MDE. \\    

\begin{tcolorbox}[%
			enhanced, 
			breakable,
			skin first=enhanced,
			skin middle=enhanced,
			skin last=enhanced,
			before upper={\parindent15pt},
			]{}

            \vspace{0.05in}

           \paragraph{\textbf{Free parameters of the model and fitting with observations.}}


The model is expressed in terms of five free parameters, $\{l_0,\beta,\epsilon_1,\eta_{\rm rh},\eta_{\rm eq}\}$.
These parameters need to be fitted to our observed universe to model a consistent evolution.
One should hence impose five conditions:
the scalar spectral index fixes $\beta$;
the amplitude of the primordial power spectrum fixes $\epsilon_1$;
entropy conservation relates the end of inflation to today and fixes $\eta_{\rm rh}$;
continuity of $\cs ^2$ at the radiation--matter transition fixes $\eta_{\rm eq}$;
and the value of the Hubble parameter today fixes $l_0$.
We now present each condition and the resulting expression in turn. \\

\begin{enumerate}
    \item \textbf{Scalar spectral index.} The constraint $\nS  - 1 = 3 - 2\nu = 4 + 2\beta$ fixes
    \begin{equation}
        \beta = -\frac{1}{2}(5-\nS )\,.
    \end{equation}

    \item \textbf{Amplitude of the primordial power spectrum.} The measured nearly scale-invariant power spectrum fixes
    \begin{equation}\label{eq:epsilon_1}
        \epsilon_1 = \frac{3^{(\nS -1)/4}}{2^{(\nS +5)/2}\,\pi^2\,\AS }\left(\frac{k_\star}{a_0\,H_0}\right)^{\nS -1}\left(\frac{a_0}{a_{\rm eq}}\right)^{(\nS -1)/4}\left(\frac{H_0}{M_{\rm Pl}}\right)^{2}\left(\frac{H_{\rm rh}}{H_0}\right)^{(5-\nS )/2}
    \end{equation}

\item \textbf{Entropy conservation.} The conservation of comoving entropy, $T\,a\,g_{*s}^{1/3} = \mathrm{const}$, relates the end of inflation to today and fixes
\begin{equation}
    \eta_{\rm rh}= \frac{3^{1/4}(\nS -3)}{2\sqrt{2}\,a_0\,H_0}\left(\frac{a_0}{a_{\rm eq}}\right)^{\!1/4}\left(\frac{H_0}{H_{\rm rh}}\right)^{\!1/2}\,,\label{eq:etareheating}
\end{equation}
    where $T_{\rm rh}$ is the reheating temperature, varying depending on the inflationary energy scale at play, $g_0 \equiv g_{*s}(T_0) = 43/11$, and $g_{\rm rh} \equiv g_{*s}(T_{\rm rh})$.

\item \textbf{Continuity of $\cs ^2$ at the radiation--matter transition.} This fixes
\begin{equation}
    \eta_{\rm eq} = \frac{2}{\sqrt{3}\,a_0\,H_0}\left(\frac{a_{\rm eq}}{a_0}\right)^{\!1/2}\left[1 + \frac{3^{3/4}\sqrt{2}\,(\nS -5)}{8}\left(\frac{a_0}{a_{\rm eq}}\right)^{\!3/4}\left(\frac{H_0}{H_{\rm rh}}\right)^{\!1/2}\right]\,.
\end{equation}
Note that the combination $\eta_{\rm eq} - \frac{\nS -5}{\nS -3}\,\eta_{\rm rh} = \frac{2}{\sqrt{3}\,a_0\,H_0}\left(\frac{a_{\rm eq}}{a_0}\right)^{\!1/2}$ is independent of the reheating temperature, as the $H_{\rm rh}$-dependent terms cancel exactly. This proves to be an important cancellation when computing the independence of the squeezing angle form the inflationary energy scale at the end of the radiation era.
\item \textbf{Hubble parameter today.} The condition $\mathcal{H}(\eta_0) = H_0\,a_0$ fixes
\begin{align}
    l_0 = \frac{4\,a_0^2\,H_0}{\sqrt{3}\,(3-\nS )}\left(\frac{a_{\rm eq}}{a_0}\right)^{\!1/2}\left[\frac{3^{1/4}(3-\nS )}{2\sqrt{2}\,a_0\,H_0}\left(\frac{a_0}{a_{\rm eq}}\right)^{\!1/4}\left(\frac{H_0}{H_{\rm rh}}\right)^{\!1/2}\right]^{\!\frac{5-\nS }{2}}\,.
\end{align}
\end{enumerate}
\end{tcolorbox}

The following Table~\ref{table_scalings} summarizes the scaling of the introduced parameters introduced with respect to the physical quantities.

\renewcommand{\arraystretch}{1.5}
\begin{table}[h!]
\centering
\begin{tabular}{cccc}
\toprule
\textbf{Quantity} &  
\textbf{$\alpha$} & \textbf{$\gamma=-\tfrac{\alpha}{\beta}$} & \textbf{$\delta=\tfrac{\alpha}{2}$} \\
\midrule
Power-law index $\boldsymbol{\beta}$ & $0$ & $0$ & $0$ \\
Scale-factor normalization $\boldsymbol{l_0}$ & $-\tfrac12(5-\nS )$ & $-1$ & $-\tfrac14(5-\nS )$ \\
Transition times $\boldsymbol{\eta_{\rm rh}},\,\boldsymbol{\eta_{\rm eq}}$ & $-1$ & $-\tfrac{2}{5-\nS }$ & $-\tfrac12$ \\
First slow-roll parameter $\boldsymbol{\epsilon_1}$ & $5-\nS $ & $2$ & $\tfrac12(5-\nS )$ \\
Hubble rate at horizon exit $\boldsymbol{H_{\rm inf}}$ & $\tfrac12(5-\nS )$ & $1$ & $\tfrac14(5-\nS )$ \\
Hubble rate at end of inflation $\boldsymbol{H_{\rm rh}}$ & $2$ & $\tfrac{4}{5-\nS }$ & $1$ \\
Scale factor at end of inflation $\boldsymbol{a_{\rm rh}}$ & $-1$ & $-\tfrac{2}{5-\nS }$ & $-\tfrac12$ \\
Expansion factor at reheating $\boldsymbol{\exp N_{\rm inf}}$ & $\tfrac12(3-\nS )$ & $\tfrac{3-\nS }{5-\nS }$ & $\tfrac14(3-\nS )$ \\
\bottomrule
\end{tabular}
\caption{Scaling of the different quantities with respect to the inflationary energy scale, with exponents defined by $Q \sim T_{\rm rh}^{\alpha}\sim H_{\rm inf}^{\gamma}\sim H_{\rm rh}^{\delta}$ and $\beta=-\tfrac12(5-\nS )$.}
\label{table_scalings}
\end{table}
\renewcommand{\arraystretch}{1.0}


\section{Super-horizon jumps at the reheating transition}

At the transition the equation of state and sound speed change from $w^{-} = -1 + 2\epsilon_1/3$, $\cs^{-} = 1$ to $w^{+} = 1/3$, $\cs^{+} = 1/\sqrt{3}$.
The leading super-horizon jumps follow directly from the matching conditions~\eqref{matching_condition_v_1}--\eqref{matching_condition_v_2}.
On super-Hubble scales $v_k'\simeq(z'/z)v_k$ because $\zeta_k'\simeq0$, so the bracket of~\eqref{matching_condition_v_2} is subleading and the condition is automatically satisfied at this order, while~\eqref{matching_condition_v_1} reduces to the continuity of $\zeta_k$,
\begin{equation}\label{eq:vk_over_z_continuous}
    \left[\frac{v_k}{z}\right]_{\pm}=0\,,
\end{equation}
at leading order $v_k$ jumps like $z$ alone.
The same holds for $v_k'\simeq(z'/z)v_k$, since $z'/z\simeq\mathcal{H}$ is continuous at the transition.
With $z^2=3M_{\rm Pl}^2(1+w)a^2/\cs^2$ and $[a]_\pm=0$,
\begin{equation}\label{eq:z_jump}
    \frac{z_+}{z_-}=\sqrt{\frac{(1+w^+)/\cs^{+2}}{(1+w^-)/\cs^{-2}}}=\sqrt{\frac{6}{\epsilon_1}}\,.
\end{equation}
In the strong-squeezing super-horizon regime, $\ee^{2r_k}\simeq|v_k'|^2/(\cs k)$ and $\phi_k\simeq\pi/2-\cs k\,|v_k|/|v_k'|$, 
so the squeezing jump follow from
the scaling of the mode functions alone,
\begin{align}
\left[r_k\right]_{\pm}\big\rvert_{\eta_{\rm rh}}
&=\log\frac{z_+}{z_-}+\frac12\log\frac{\cs^-}{\cs^+}
=\frac{1}{2}\log\frac{6\sqrt{3}}{\epsilon_1}\,,\label{eq:rjump_leading}
\end{align}
The angle jump does not enter the coherence budget; we nonetheless present its scaling for pedagogical purposes,
\begin{align}
\frac{\pi/2-\phi_{k+}}{\pi/2-\phi_{k-}}\bigg\rvert_{\eta_{\rm rh}}
&=\frac{\cs^+}{\cs^-}=\frac{1}{\sqrt3}\,,\label{eq:deltajump_leading}
\end{align}
Since the
phase variance is a function of the squeezing amplitude alone,
$-\log\Delta\theta_k\simeq r_k-\tfrac12\log(2\ln2)$, its jump follows without
further computation,
\begin{equation}\label{eq:logdtheta_jump}
\left[-\log\Delta\theta_k\right]_{\pm}\big\rvert_{\eta_{\rm rh}}
=\left[r_k\right]_{\pm}
=\frac{1}{2}\log\frac{6\sqrt{3}}{\epsilon_1}\,,
\end{equation}
controlled by $\epsilon_1$ alone; equivalently,
$\Delta\theta_k\,z/\sqrt{\cs}\propto\Delta\theta_k\sqrt{(1+w)/\cs^{3}}$ is
continuous at the transition.


\section{Scaling of the squeezing and the phase variance}

\paragraph{Inflation.} Substituting the mode function solution~\eqref{eq:mode_inflation} in
the power spectra~\eqref{eq:powers_pectra}, themselves inserted in the squeezing
parameters~\eqref{eq:squeezing}, we obtain on super-Hubble scales $- k \eta \ll 1$,
\begin{equation}\label{eq:scaling_r_inf}
r_k\simeq-\frac{5-n_s}{2}\log(-k\eta)\simeq\frac{5-n_s}{3-n_s}\,N_{\rm inf}\,,
\qquad
\phi_k\simeq\frac{\pi}{2}-e^{-2N_{\rm inf}/(3-n_s)}\,,
\end{equation}
i.e.\ the amplitude grows linearly in the number of $e$-folds while the angle
tends exponentially towards the momentum direction $\pi/2$ as the mode
freezes (the latter does not enter the coherence growth and is reported for
completeness). The phase variance thus sharpens as
\begin{equation}\label{eq:scaling_logdtheta_inf}
-\log\Delta\theta_k\simeq\frac{5-n_s}{3-n_s}\,N_{\rm inf}\,.
\end{equation}
For the exact de Sitter case
$n_s=1$, $-\log\Delta\theta_k\simeq r_k\simeq2N_{\rm inf}$.

\paragraph{Radiation.}
Because the scale factor~\eqref{eq:scale_factor_RDE} is linear in conformal
time, $z''/z=0$ and the mode functions obey a free oscillator with constant
frequency $\omega=\cs k$. With no ongoing particle production, the squeezing
amplitude, and with it the phase variance, is \emph{frozen} at all scales
throughout the radiative evolution, its constant value set by the
inflationary accumulation~\eqref{eq:scaling_r_inf} at reheating plus the
jump~\eqref{eq:rjump_leading},
\begin{equation}\label{eq:logdtheta_anatomy}
-\log\Delta\theta_k\big\rvert_{\rm rad}
=\underbrace{\frac{5-n_s}{3-n_s}N_{\rm inf}}_{\text{inflationary build-up}}
+\underbrace{\frac12\log\frac{6\sqrt3}{\epsilon_1}}_{\text{reheating jump}}
-\frac12\log(2\ln2)\,.
\end{equation}
The squeezing angle, by contrast, keeps running: it drifts at the constant
rate set by the matching~\eqref{eq:deltajump_leading}, with
$\pi/2-\phi_k\propto a$, reaching at equality the $H_{\rm rh}$-independent
value
\begin{align}\label{eq:phifin}
    \phi_k^{\rm rad}(\eta_{\rm eq})&\simeq\frac{\pi}{2}-\frac{5-n_s }{3-n_s }\frac{-k\eta_{\rm rh}}{\sqrt{3}}
    -\frac{k\eta_{\rm eq}}{\sqrt{3}}\,\pmod\pi\\
    &\simeq\frac{\pi}{2}-\frac{2k}{3\,a_0\,H_0}\sqrt{\frac{a_{\rm eq}}{a_0}}\,\pmod\pi\,.
\end{align}

\paragraph{Arbitrary epoch.}
More generically, for an arbitrary epoch with equation of state parameter $w$, the scaling of the phase variance on super-Hubble scales is given by
\begin{equation}\label{eq:Deltatheta_arbitraryepoch}
-\log\Delta\theta_k\simeq \frac{1-3w}{2}N_{\rm ef}\,,
\end{equation}
where $N_{\rm ef}$ is the number of e-folds since the beginning of the epoch. This is because $v_k\propto z \propto a\propto \eta^{2/(1+3w)}$ and $\Delta\theta_k\propto |v_k^{\prime}|\propto a^{(1-3w)/2}$. Hence for a strict presureless dominated era, $-\log\Delta\theta_k\simeq 1/2 N_{\rm ef}$.


\section{Independence from the inflationary energy scale.}

We can now assemble the variance and track the origin of its scale-independence term by term.
Evaluating~\eqref{eq:logdtheta_anatomy} at the radiation--matter transition for the inflationary scalings~\eqref{eq:scaling_r_inf} together with the jump~\eqref{eq:logdtheta_jump} yields
\begin{equation}\label{eq:logdtheta_anatomy_supp}
    -\log\Delta\theta_k\big\rvert_{\eta_{\rm eq}}
    =\frac{5-\nS }{3-\nS }N_{\rm inf}
    +\frac12\log\frac{6}{\epsilon_1}
    +\mathcal{O}(1)\,,
\end{equation}
where $\mathcal{O}(1)$ is a pure number that does not depend on the inflationary energy scale. This expression is Eq.~\eqref{eq:logdtheta_scaling} in the main text when specialized to the de Sitter case ($\nS =1$).
From the scalings in Table~\ref{table_scalings}, $N_{\rm inf}\sim\tfrac{3-\nS }{4}\log H_{\rm rh}$ which gives $\tfrac{5-\nS }{3-\nS }N_{\rm inf}=\tfrac{5-\nS }{4}\log H_{\rm rh}+\text{const}$, under which~\eqref{eq:logdtheta_anatomy_supp} reduces to~\eqref{eq:logdtheta_cancel}, namely
\begin{equation}\label{eq:invariance_slowroll}
    -\log\Delta\theta_k\big\rvert_{\eta_{\rm eq}}=\frac{1}{2}\log \frac{H_{\rm inf}^2}{\epsilon_1}+{\rm const}\,,
\end{equation}
using $H_{\rm inf}^2\sim H_{\rm rh}^{(5-\nS )/2}$.
Equation~\eqref{eq:invariance_slowroll} is thus independent of the inflationary energy scale, generalizing the main-text proof to include slow-roll corrections.

\end{widetext}

\end{document}